\documentclass[letterpaper,twocolumn,10pt]{article}
\usepackage{usenix2019_v3}

\usepackage{tikz}
\usepackage{amsmath}
\usepackage{graphicx}
\usepackage{multirow,tabularx}
\usepackage{tablefootnote}
\usepackage{booktabs}

\begin{document}

\date{}

\title{\Large \bf Automated Mapping of CVE Vulnerability Records to MITRE CWE Weaknesses}

\author{
{\rm \large Ashraf Haddad $^\dagger$ } 
\and
{\rm \large Najwa Aaraj $^{\dagger \ast}$ }
\and
{\rm \large Preslav Nakov $^{\dagger}$}
\and
{\rm \large Septimiu Fabian Mare $^\ast$}
\and
{\rm} \\
{\small $^\dagger$ MBZUAI- Mohammed Bin Zayed University for Artificial Intelligence, Abu Dhabi, UAE} \\
{\small $^\ast$ TII - Technology Innovation Institute, Abu Dhabi , UAE }
} 
\maketitle

\begin{abstract}
In recent years, a proliferation of cyber-security threats and diversity has been on the rise culminating in an increase in their reporting and analysis. To counter that, many non-profit organizations have emerged in this domain, such as MITRE and OSWAP, which have been actively tracking vulnerabilities, and publishing defense recommendations in standardized formats. As producing data in such formats manually is very time-consuming, there have been some proposals to automate the process. Unfortunately, a major obstacle to adopting supervised machine learning for this problem has been the lack of publicly available specialized datasets. Here, we aim to bridge this gap. In particular, we focus on mapping CVE records into MITRE CWE Weaknesses, and we release to the research community a manually annotated dataset of 4,012 records for this task. With a human-in-the-loop framework in mind, we approach the problem as a ranking task and aim to incorporate reinforced learning to make use of the human feedback in future work. Our experimental results using fine-tuned deep learning models, namely Sentence-BERT and rankT5, show sizable performance gains over BM25, BERT, and RoBERTa, which demonstrates the need for an architecture capable of good semantic understanding for this task.

\end{abstract}

\section{Introduction}

\begin{figure}
    \centering
    \includegraphics[width=\linewidth]{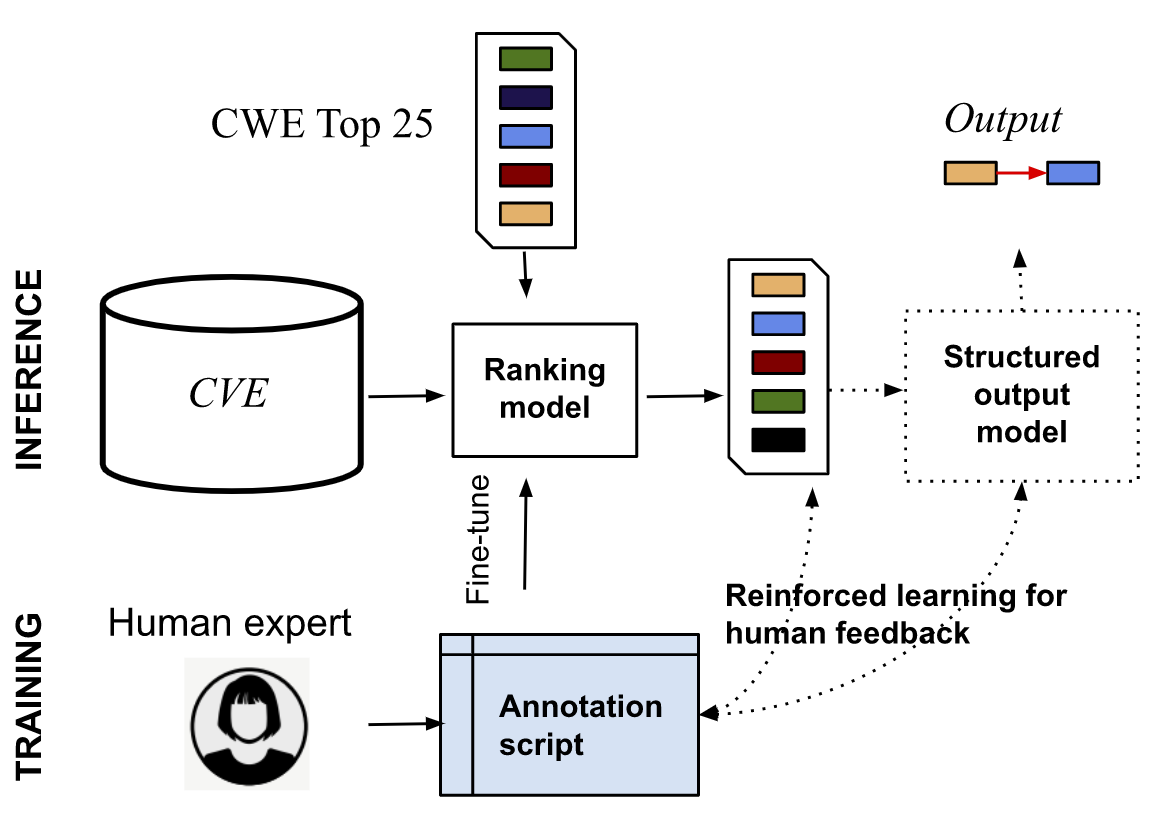}
    \caption{Our framework for mapping vulnerability records to structured output, with human-in-the-loop feedback.}
    \label{fig:pipeline}
\end{figure}

\begin{figure*}
    
    \centering
    \includegraphics[width=\linewidth]{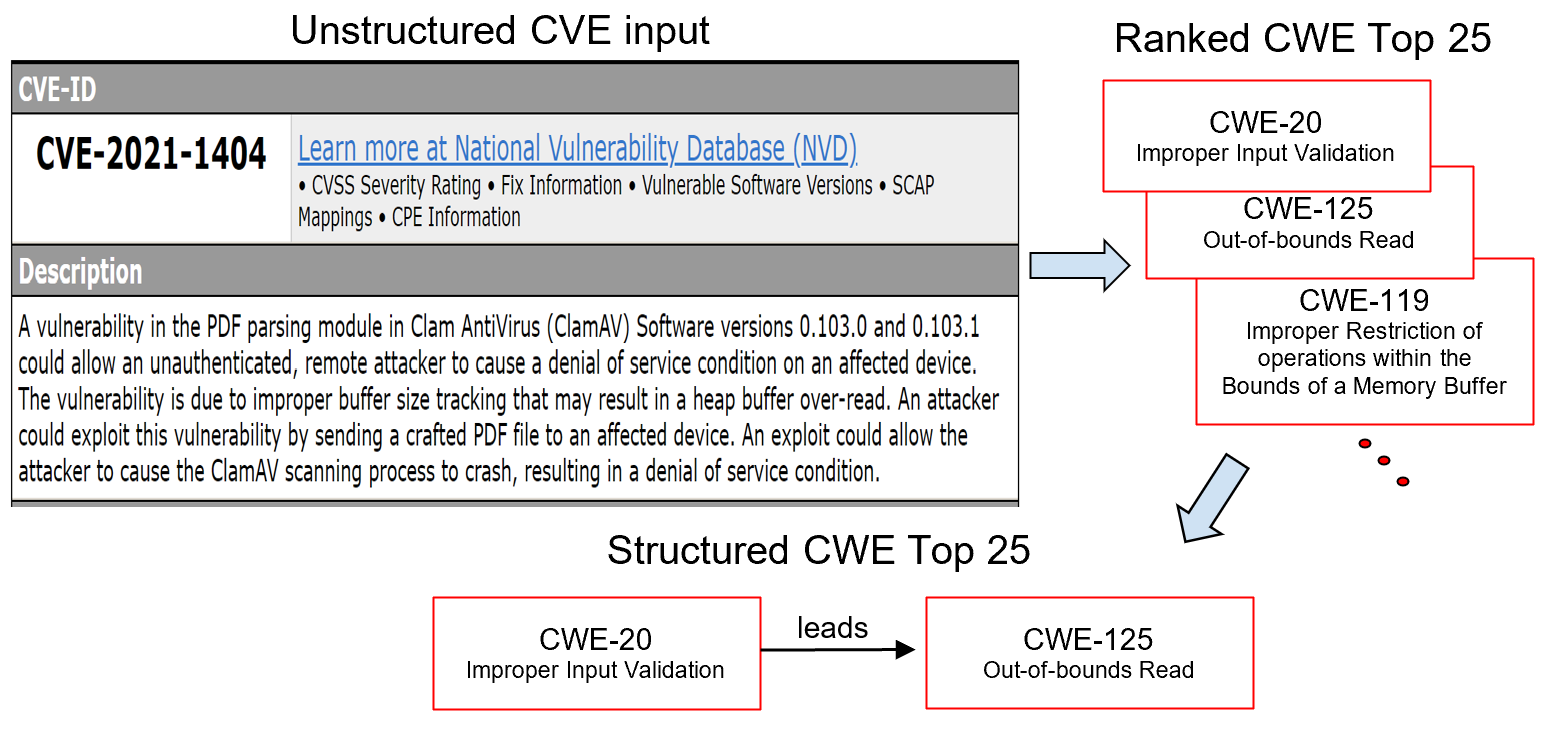}
    \caption{\label{fig:medium} Our task in focus: mapping unstructured input to MITRE CWE Top 25 Weaknesses.}
\end{figure*}

Cyber-security is a major technological concern demonstrated by the multitude of threats constantly mutating and being reported~\cite{CVE} across many systems, including  operating systems, Internet-of-Things(IoT), physical networks, and other tiered technologies touching every aspect of the economy and society. For example, a 2021 OWASP top 10~\cite{OWASP} report shows that three out of the top 10 most critical security vulnerabilities were not as critical or present a few years earlier.

Cyber Threat Intelligence (CTI) reporting aims at identifying and organizing the information in logged vulnerabilities. Both non-profit organizations and private institutions are keen on producing such reports to end-users, thus educating designers and developers of systems on tackling such weaknesses.

The types of CTI depends on the audience and on how technical the analysis ought to be. Some reports are high-level and aimed at providing non-technical audience with strategic and impact analysis, while others are technical, including details about IoCs (Indicators of Compromise) and the version of the specific software/hardware affected. Regardless of the technical depth, efficiently sharing CTI reports requires a common standardization that is widely adopted. The MITRE\footnote{\url{https://www.mitre.org/}} Top 25 Common Weakness Enumeration(CWE)~\cite{CWE} is one known community-developed weakness types describing the most impactful vulnerabilities. These vulnerabilities, such as the ones reported by the Open Source Intelligence (OSINT) community on the National Vulnerability Database (NVD) CVE~\cite{CVE}, require cyber-security experts' analysis to deduce the underlying weaknesses. This is a tedious and  time-consuming task, yet a prime candidate for AI (Artificial Intelligence) tools to assist in. Thus, the motivation behind our work here is to map reported CVE vulnerabilities to the underlying CWE weaknesses, which is a vital step in producing CTI reports. 

In the spirit of open-source research, our focus is on leveraging the records present in the publicly available NVD CVE and utilize them for the task of standardization using Natural Language Processing (NLP), thus benefiting the cyber-security and the AI communities at large. This sets apart our approach to the problem where the team's effort went into building a dataset along with the tools needed to maintain it, considering the ever-changing cyber-security domain. Moreover, we validated state-of-the-art ranking models on that dataset. Our best results are achieved when using transformer-based architectures in two configurations: a semantic similarity matching setup and a sequence-to-sequence ranking setup to retrieve the most appropriate CWE type describing the issue. Our contributions can be summarized as follows:
\begin{itemize}
    \item \textbf{A new problem formulation:} using AI to assist human annotators to streamline the analysis and the labeling of logged vulnerabilities in the NVD dataset, as shown in Figure~\ref{fig:pipeline}. The model output can leverage the reinforced learning from annotator's feedback~\cite{RLHF}, which can be explored in future work.  
    \item \textbf{A new dataset:} The above formulation cannot be implemented without first creating a suitable dataset. We release our dataset
   \footnote{ \url{https://github.com/ahadda5/annotate\_cve}} .
    \item \textbf{Experimental results.} We approached the task as a ranking problem, using text similarity  SBERT~\cite{Reimers-SBERT}, optimized for sentences (vs. the document level) and the recently released ranking T5~\cite{monoT5} model on the dataset. Our results were compared to the more general-built models BERT~\cite{BERT} and RoBERTa~\cite{Liu-etal-Roberta}. BM25~\cite{BM25} was used as the baseline model. Moreover, we used T5 as Seq2Seq generation~\cite{seq2seq} model as well as a ranker.
\end{itemize}

Figure \ref{fig:medium} highlights the aim of our work at large which is inline with MITRE's vision on the standardization of CVE records, an extremely valuable task to the community \footnote{  A call-to-action by MITRE using ATT\&CK framework \scriptsize \url{https://medium.com/mitre-engenuity/cve-mitre-att-ck-to-understand-vulnerability-impact-c40165111bf7}} .


The rest of this paper is organized as follows: Section~\ref{sec:2} provides a
summary of related work. Section~\ref{sec:3} briefly summarizes the MITRE weaknesses types and how they are used. Section~\ref{sec:4} explains the rationale behind annotating and maintaining the released cyber-security AI dataset. Section~\ref{sec:meth} introduces our methodology. Section~\ref{sec:expNeval} describes our experimentation and discusses the results. Finally, Section~\ref{sec:concNFut} concludes and points to possible directions for future work.

\section{Related Work}
\label{sec:2}
The research topic revolving around the application of NLP to CTI has peaked interest in recent years. This is driven by common reasons related to the rise of AI: the sheer availability of data, despite not always AI-ready, and the advances in algorithms and their execution. NLP is necessary to extract the relevant information and the intrinsic relationships between entities. The type of research in this field is divided into two area based on the model design. Some rely on traditional Machine learning (ML) methods such as Support Vector Machines (SVM) \cite{Luke,SVM1}, or probabilistic reasoning \cite{RuleReasoning}, while others use deep learning models such as LSTM (Long short-term memory) and transformers \cite{GasmiLSTM,TCEnet}. Learning representation at such a high dimensionality of features with a dynamic corpus, like that of cyber-security, is better suited with deep learning methods, especially with access to High Performance Computing (HPC) machines. 

Zhou et al.~\cite{IOC-CRF} used an LSTM with a Conditional Random Fields (CRF) layer for sequence labeling focused solely on IoC identification. While high accuracy was achieved this is only a portion of the task at hand since standardization requires higher-order descriptions beyond IoCs like \emph{Missing Authorization} or \emph{Server-side Request Forgery}. Moreover, CRF being a discriminative label sequencing model requires feature engineering, which needs over-specified manually designed features in comparison to, say, simply training a transformer model on new data and labels. Zhu and Dumitras~\cite{FeatureSmith} recognized the challenge with feature engineering and proposed parsing sentences using a semantic network model, which aims at mirroring the human process of reasoning in manual feature design. Their model assigns weights to important concepts and relation types by calculating their semantic similarity. The concepts are first extracted as a behavior tuple of subject, verb and object through a dependency parse, then weighed based on maximum syntactical score of the verb and noun phrase. The semantic network then ranks a pool of potential features to determine the most relevant in describing the vulnerability. Zhu and Dumitras~\cite{ChainSmith}, continued their work by approaching the IoC identification and classification with an elaborate pipeline model. The web-crawled reports underwent a syntactic and semantic parse before being processed through named entity recognition (NER) on possible IoC strings which was finally processed through a binary classifier. All the above work only tackles NER of specifics like IoC and thus it does not serve our goal of automated mapping. 

Information Retrieval (IR) using BM25 was previously used on a large corpus of Symantec threat reports \cite{TTPDrillHusari}. Part-Of-Speech (POS) tagging and BM25 with TF.IDF weighting identified and characterized malicious actions from the CTI report. To overcome the massive information presented in those documents, an SVM classifier was trained to filter out the information within the document  with little relevance. The sanitized content was then parsed though POS tags to find potential threat actions using subject, verb, and object tuples. The results were then scored against the ATT\&CK patterns to find the best matches. Ayoade et al.~\cite{Extend} instead considered classifying threats into tactics, techniques, and killChains independently and applied a bias correction due to  ``limited labeled training data.'' This bias was mitigated through the use of covariant shift like kernel mean matching (KMM). The limited amount of training data is an observation we agree with and is one of the main motivations to annotate and release our dataset. 

Legoy et al.~\cite{Legoy} defined the problem as a multi-label text classification extracting ATT\&CK TTPs (tactics, techniques, and procedures) from the limited annotated open-source reports available. They thoroughly compared a number of traditional classifiers including Logistic Regression, Na\"{i}ve Bayes, K-nearest Neighbors, AdaBoost Decision Trees, etc. They trained and evaluated independently for techniques vs. tactics. Moreover, since techniques could belong to one or more tactics, the associated tactic score could be used with the existing technique score to boost an ensemble confidence score for the techniques. For finding relationships between techniques (similar to the bottom part of Figure \ref{fig:medium}), the paper uses statistical methods computing the joint probability of pairs of techniques like the Rare association rule or the Steiner tree association rules. However, they did not achieve good results with these methods as they could ``probably fit better in a hierarchical environment with known conditional probabilities.''

In terms of deep learning, a number of papers stand out. Gasmi et al. used LSTMs to for NER and relation extraction (RE) \cite{GasmiLSTM}. The first task was performed using a tier of a word-embedding layer, followed by a bi-directional LSTM and a final CRF layer to perform NER labels (adapted from \cite{LSTMLample}). For the RE, the LSTM with the Shortest Dependency Path (SDP) \cite{LSTM-SDP} and the LSTM on Sequences and Tree Structures \cite{LSTM-STS} were explored. The clear shortcoming of the paper is the usage of machine-labeled data, which is not precise to start with. For the NER labeled data, Stucco project \cite{Stucco} was leveraged, which does not use standards like CWE or ATT\&CK TTP labels. Furthermore, a bootstrapping algorithm \cite{RE_Jones} was used to generate label relationships for the training of the RE task.  
Encouragingly, the recently released TCENet \cite{TCEnet} achieved high accuracy on manually annotated security reports. The model finds and classifies TTP descriptions and further extracts element features in the description, at the sentence vs. the document level of the report to enhance the accuracy. The work challenges previous work, which relied on subject-verb-object tuples \cite{TTPDrillHusari,FeatureSmith} stating that not all TTPs can simply be modeled as such. Moreover, previous models used the static TTP names only and did not rely on TTP descriptions. A key motivation we exploit here when calculating SBERT or BM25 similarity measures explicitly, or implicitly through BERT, RoBERTa or T5, is to use both descriptions and names. Despite annotating many reports in the above work, they are not accessible to the public (only few examples were released); in contrast, we release our dataset in full.

Finally, the research by Marchiori et al. \cite{STIXnet} divided the problem into a pipeline of entity extraction followed by relation extraction. The entity extraction, itself, is a combination of regular expression for IoCs, rule-based methods using a large knowledge-based dataset, dependency parsing and NER deep learning. Relation extraction is done using POS tagging and dependency parsing to build a graph for each sentence and to retrieve relations by looking at the shortest paths between entities. These RE results are compared to a second approach where a transformer model computed embeddings of the sentences and their similarity with ATT\&CK labels. Only 50 reports were annotated for training, which is insufficient to train a deep learning model, considering there are more than a hundred TTP labels. 

It is clear that previous research had two major drawbacks. First, an annotated open-source dataset is not available or maintained. We aim to bridge this gap. Second, aside from \cite{STIXnet,TCEnet}, previous research ran models at the document and not at the sentence level and only focused on the actual names of the weaknesses, ignoring their description. This attenuates important underlying features relevant to the task. Moreover, a lot of previous research focused on the automatic mapping as either a classification task or as an entity and relation extraction task, when the very nature of cyber-security experts' analysis is subjective and their labels are not always matching. Ergo, a ranking approach is more appropriate. Finally, to the best of our knowledge, no current research is leveraging the importance of advances in human feedback model training, but that is partly because annotation and an AI-compatible dataset was not invested in. 
\section{2022 CWE Top 25 Weaknesses}
\label{sec:3}
\begin{figure*}[h]
    \centering
    \includegraphics[width=\linewidth]{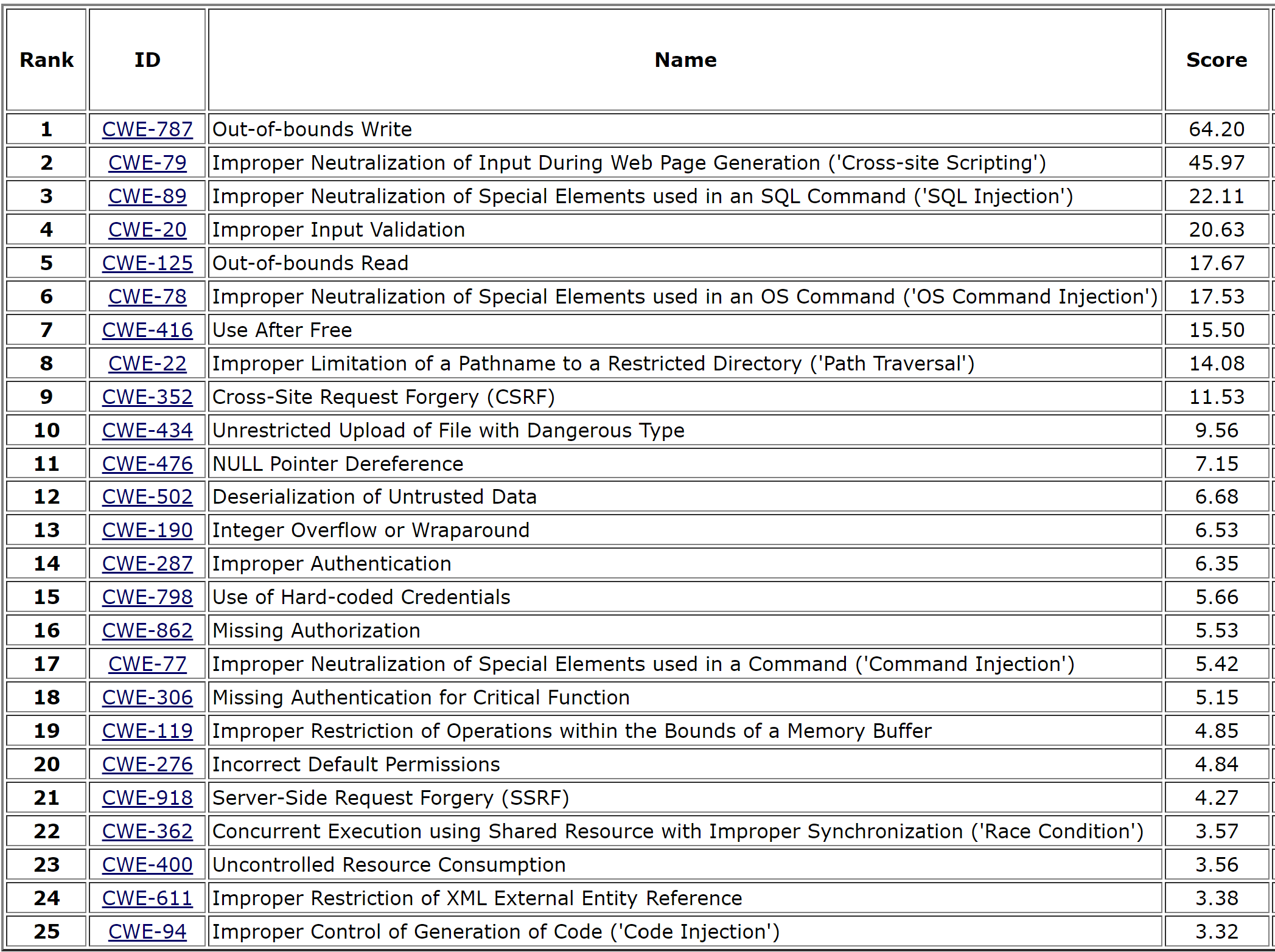}
    \caption{The MITRE 2022 CWE Top 25 Weaknesses \cite{CWE}. The rank is based on the CVSS score shown.}
    \label{fig:top25}
\end{figure*}

The CWE Weaknesses list~\cite{CWE}  focuses on application security and describes the common weaknesses exploited by attackers.  Many cyber-security professionals find this list to be a practical and a convenient resource to help mitigate risks. It enables application threat modeling to understand the problems in the design of the software and/or hardware. We chose CWE due to its familiarity with the group of available annotators. Moreover, in order to focus our efforts, we chose the Top 25 weaknesses. The danger of each weakness is scored based on the Common Vulnerability Scoring System (CVSS) metric adopted by NIST~\cite{CVSS}. The list is dynamic and changes yearly based on the analysis of the most impactful CVE vulnerabilities. The entire set for 2022 covers 37,899 CVE records for the previous two years. Those records would have been approved and triaged into CWE weaknesses through human analysts. Future work should focus on enabling AI to assist human analysts  (as shown in Figure~\ref{fig:pipeline}). Moreover, what we use for AI training are the CWE Top 25  names, descriptions, and extended descriptions to deduce the relevance of a weakness or a group of weaknesses given the logged vulnerabilities. Figure~\ref{fig:top25} shows the 2022 top 25 CWE.  It is important to note that our annotation approach can be adapted to MITRE ATT\&CK framework \cite{MITRE} by utilizing its TTP (tactics, techniques, and procedures) instead of CWE. Once an ATT\&CK AI dataset is established the same methodology, as in Section~\ref{sec:meth}, can be used. 

\section{Dataset}
\label{sec:4}
The need for a dataset in this domain is crucial akin to the importance of, say ,ImageNet~\cite{imagenet} to Computer Vision tasks or the GLUE benchmark (General Language Understanding Evaluation) ~\cite{glue} to NLP tasks. While many papers claimed this exercise was done for MITRE ATT\&CK, either fragments of data were released or nothing was published at all. However, as previously stated, due to the ever-changing field of cyber-security, a robust approach is needed to account for threats and their severity scores. Therefore, as part of this paper's contribution, we created a separate code \footnote{\url{https://github.com/ahadda5/annotate\_cve}\label{annotate_cve}}  
repository to annotate and maintain such a dataset. 

The NVD is the starting point of our dataset. We review and annotate it to create an AI-compatible dataset to the task of ranking and label relationship. The annotation process is influenced by the prospective of the annotator reasoning. As noticeable in the NVD itself, one CVE record could have multiple CWE labels; it could have one CWE label by one annotator and different label(s) by another annotator. Also, some records might be mis-categorized as shown in Table~\ref{tab:NVDvsOurs}. Moreover, in some CVE descriptions, NVD annotators could choose multiple labels, when in fact it is clear that one led to the other, a task which current NLP methods can deduce with relatively high accuracy~\cite{depParsing}. As commonly known in the AI community, any model is as good as the data it is trained on. It thus merits the effort to review NVD's annotated CVE records, to choose the most relevant CWE labels (the focus is on the 2022 CWE Top 25) and to infer the dependency relations between labels (as in Figure~\ref{fig:medium}), if found in the log description, rather than just assigning multiple labels. While our review process agrees with the majority of NVD's CWE labels ($\approx 77\%$) which make their way untouched to the released dataset, some required choosing more appropriate labels or noting the dependencies between them. Table~\ref{tab:NVDvsOurs} shows such examples.

\begin{table*}[]
\centering
\begin{tabular}{p{2.1cm}p{5.8cm}p{2.5cm}p{2.4cm}p{2.4cm}}
\toprule
\textbf{NVD CVE-ID} & \textbf{CVE Description} & \textbf{NVD Labels} & \textbf{Our Labels} & \textbf{Remark}\\
\midrule
\footnotesize{CVE-2021-30129} & A \footnotesize{vulnerability in sshd-core of Apache Mina SSHD allows an attacker to overflow the server causing an OutOfMemory error. This issue affects the SFTP and port forwarding features of Apache Mina SSHD version 2.0.0 and later versions. It was addressed in Apache Mina SSHD 2.7.0}  &  \footnotesize{CWE-772 (Missing Release of Resources after Effective Lifetime)}  &  \footnotesize{CWE-787 (Out-of-bounds Write)} &  \footnotesize{More relevant CWE label}\\
\midrule
\footnotesize{CVE-2022-36796} & \footnotesize{Cross-Site Request Forgery (CSRF) vulnerability leading to Stored Cross-Site Scripting (XSS) in CallRail, Inc. CallRail Phone Call Tracking plugin <= 0.4.9 at WordPress}  &  \footnotesize{CWE-352 (Cross-Site Request Forgery) and CWE-79 (Cross-site Scripting)}  &  \footnotesize{CWE-352 leads to CWE-79} & \footnotesize{Causation present or inferred in description} \\
\midrule
\footnotesize{CVE-2020-13821} & \footnotesize{A remote code execution vulnerability exists when the Windows font library improperly handles specially crafted fonts.For all systems except Windows 10, an attacker who successfully exploited the vulnerability could execute code remotely, aka 'Windows Font Library Remote Code Execution Vulnerability'}  &  \footnotesize{CWE-787 (Out-of-bounds Write)} & \footnotesize{CWE-20 (Improper Input Validation) leads to CWE-94 (Improper Control of Generation of Code)} & \footnotesize{More relevant CWE labels and causation present} \\
\midrule
\footnotesize{CVE-2020-11854} & \footnotesize{Arbitrary code execution vulnerability in Operation bridge Manager, Application Performance Management and Operations Bridge (containerized) vulnerability in Micro Focus products products Operation Bridge Manager, Operation Bridge (containerized) and Application Performance Management. The vulnerability affects: 1.) Operation Bridge Manager versions 2020.05, 2019.11, 2019.05, 2018.11, 2018.05, 10.63,10.62, 10.61, 10.60, 10.12, 10.11, 10.10 and all earlier versions. 2.) Operations Bridge (containerized) 2020.05, 2019.08, 2019.05, 2018.11, 2018.08, 2018.05. 2018.02 and 2017.11. 3.) Application Performance Management versions 9,51, 9.50 and 9.40 with uCMDB 10.33 CUP 3. The vulnerability could allow Arbitrary code execution.}  &  \footnotesize{CWE-787 (Out-of-bounds Write)} & \footnotesize{CWE-94 (Improper Control of Generation of Code)} & \footnotesize{More relevant CWE label} \\
\midrule
\footnotesize{CVE-2020-15158} & \footnotesize{In libIEC61850 before version 1.4.3, when a message with COTP message length field with value < 4 is received an integer underflow will happen leading to heap buffer overflow. This can cause an application crash or on some platforms even the execution of remote code. If your application is used in open networks or there are untrusted nodes in the network it is highly recommend to apply the patch. This was patched with commit 033ab5b. Users of version 1.4.x should upgrade to version 1.4.3 when available. As a workaround changes of commit 033ab5b can be applied to older versions.}  &  \footnotesize{CWE-191 (Integer Underflow), CWE-119 (Improper Restriction of Operations within the Bounds of a Memory Buffer) and CWE-122 (Heap-based Buffer Overflow)} & \footnotesize{CWE-191 leads to CWE-122 leads to CWE-94} & \footnotesize{More relevant CWE labels and causation present} \\
\midrule
\footnotesize{CVE-2022-22243} & \footnotesize{An XPath Injection vulnerability due to Improper Input Validation in the J-Web component of Juniper Networks Junos OS allows an authenticated attacker to add an XPath command to the XPath stream, which may allow chaining to other unspecified vulnerabilities, leading to a partial loss of confidentiality. This issue affects Juniper Networks Junos OS: all versions prior to 19.1R3-S9; 19.2 versions prior to 19.2R3-S6; 19.3 versions prior to 19.3R3-S7; 19.4 versions prior to 19.4R2-S7, 19.4R3-S8; 20.1 versions prior to 20.1R3-S5; 20.2 versions prior to 20.2R3-S5; 20.3 versions prior to 20.3R3-S5; 20.4 versions prior to 20.4R3-S4; 21.1 versions prior to 21.1R3-S2; 21.2 versions prior to 21.2R3-S1; 21.3 versions prior to 21.3R2-S2, 21.3R3; 21.4 versions prior to 21.4R1-S2, 21.4R2-S1, 21.4R3; 22.1 versions prior to 22.1R1-S1, 22.1R2.}  &  \footnotesize{CWE-91 (XML Blind XPath Injection) and CWE-20 (Improper Input Validation)} & \footnotesize{CWE-20 leads to CWE91} & \footnotesize{Causation present or inferred in description} \\

\bottomrule
\end{tabular}
\caption{\label{tab:NVDvsOurs}
Examples of mismatched NVD labels and our annotation philosophy relevant to a structured output.}
\end{table*}

\begin{table}
\centering

\begin{tabular}{llc}
\toprule
\textbf{} & \textbf{CVE-ID} & \textbf{Labels} \\
\cline{2-3}
& CVE-2020-12960 & 4 \\
& CVE-2020-13959 & 2-25 \\
& CVE-2021-44224 & 11-21 \\
 \multicolumn{1}{c}{\multirow{-4}{*}{\textit{\textbf{Examples }}}} & CVE-2021-44042 & 17 \\
\bottomrule
\\
\end{tabular}

\begin{tabular}{cccc}
\toprule  
 & \bf Total & \bf Single Label & \bf Causation Label \\
\cline{2-4}
\textit{\textbf{Count }} & 4,012& 3,605 & 407 \\
\bottomrule

\end{tabular}
\caption{\label{tab:cveCweDBsample}
Example and statistics about our dataset.}
\end{table}

\begin{table*}[tbh]
\centering
\begin{tabular}{lp{14cm}r}
\toprule
\textbf{Id} & \textbf{Label} & \textbf{Count}  \\
\midrule
CWE-787 & Out-of-bounds Write & 261 \\
CWE-79 & Improper Neutralization of Input During Web Page Generation (Cross-site Scripting)  & 626\\
CWE-89 & Improper Neutralization of Special Elements used in an SQL Command  (SQL Injection) & 301  \\
CWE-20 & Improper Input Validation & 173 \\
CWE-125 & Out-of-bounds Read & 100  \\
CWE-78 & Improper Neutralization of Special Elements used in an OS Command (OS Command Injection) & 47  \\
CWE-416 & Use After Free &  35 \\
CWE-22 & Improper Limitation of a Pathname to a Restricted Directory (Path Traversal) & 137 \\
CWE-352 & Cross-Site Request Forgery (CSRF) & 92 \\
CWE-434 & Unrestricted Upload of File with Dangerous Type & 78 \\
CWE-476 & NULL Pointer Dereference & 56 \\
CWE-502 & Deserialization of Untrusted Data & 51 \\
CWE-190 & Integer Overflow or Wraparound & 39  \\
CWE-287 & Improper Authentication &  404  \\
CWE-798 & Use of Hard-coded Credentials & 92 \\
CWE-862 & Missing Authorization & 387  \\
CWE-77 & Improper Neutralization of Special Elements used in a Command (Command Injection) & 147 \\
 CWE-119 & Improper Restriction of Operations within the Bounds of a Memory Buffer & 148 \\
 CWE-276 & Incorrect Default Permissions & 78  \\
 CWE-918 & Server-Side Request Forgery (SSRF) & 86 \\
 CWE-362 & Concurrent Execution using Shared Resource with Improper Synchronization (Race Condition) & 47  \\
 CWE-400 & Uncontrolled Resource Consumption & 122 \\
 CWE-611 & Improper Restriction of XML External Entity Reference & 41 \\
 CWE-94 & Improper Control of Generation of Code (Code Injection) & 57 \\
\bottomrule
\end{tabular}
\caption{\label{DBCount}
Breakdown of single-label records in our dataset for the ranking task.}
\end{table*}

\begin{figure*}[ht]
    \includegraphics[width=\linewidth]{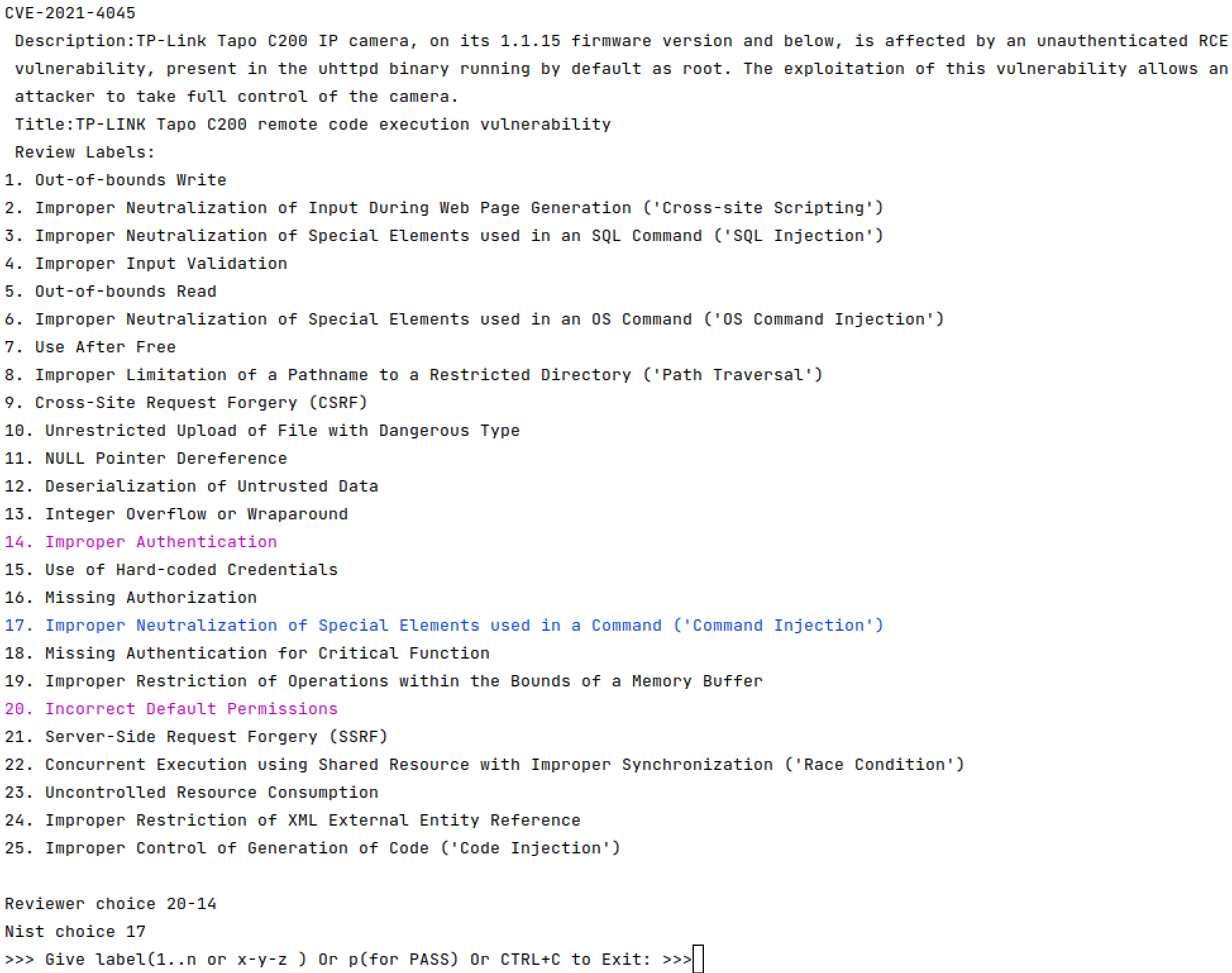}
    \caption{Annotator script for the review process. The prompt presents labels 20-14(20 leads to 14) chosen by another annotator and NIST NVD label 17. }
    \label{fig:prompt}
\end{figure*}

The process ran as such: 
\begin{enumerate}
    \item Publicly available CVE records \footnote{\url{https://github.com/CVEProject/cvelist}} corresponding to the years of the 2022 CWE Top 25 were retrieved. The data contains a number of metadata attributes for each record including external URLs related to the issue, assigner, and reporter details, among other details.  
    \item The records are filtered based on their accepted state by MITRE. Three pieces of attributes are retrieved from the data, namely the \textit{description}, the \textit{title}, and the \textit{id}.  The id is a unique expression of the format CVE-YYYY-NNNN used as a data structure index throughout the code. (YYYY is the year and NNNN are unique digits).
    \item An Information retrieval (IR) method is used to match the CVE records with the target CWE mapping increasing the likelihood of choosing records with those target labels. The goal is to narrow the annotation pool of 37,899 CVE records and focus on records which will more likely fall under the 2022 CWE Top 25 labels. The used IR model is TF.IDF (Term Frequency Inverse Document Frequency), which returns a sparse matrix of the frequency of words used for scoring the most likely weaknesses. The corpus of the CVE records is first prepossessed for lowercase and then stopwords (common words e.g. \emph{the}, \emph{and}, \emph{a}) are removed.  In essence, this step is matching the most frequent words between the 25 CWE name and description to that of the CVE description and its title. In reality, some retrieved records cannot be categorized under the CWE Top 25 and must be skipped. As shown in the Results Section~\ref{sec:expNeval}, classical IR methods are unreliable at this task and the purpose of this step is to only increase the likelihood of records with possible 2022 CWE Top 25 labels for the sole purpose of creating a dataset.  \footnote{Step 3 could be skipped, but we wanted to focus our efforts on a few thousand records to annotate}
    \item Web-crawl NVD website \footnote{Of the format:  https://nvd.nist.gov/vuln/detail/CVE-YYYY-NNNN} corresponding to the narrowed list from the step above.  This returns the CWE label(s) annotated by NIST and/or 3$^{rd}$ party vendors. 
    \item  The list of CVEs to annotate is divided between three annotators with no overlap in indexes. Each annotator will run the annotation script~$^{\ref{annotate_cve}}$ as shown in Figure~\ref{fig:prompt}. The prompt allows to either enter another label if the annotator disagrees with NVD exiting label (or labels already chosen by another annotator), skip the record if they agree, or type a causation group of labels if the CVE description merits that. Once the first round of annotation is done, the divided list is rotated and reviewed by a different annotator. If two annotators agree on a record, it is deemed fit for the final training/validation set, otherwise a third annotator is asked to review those records and finalize the outcome.
\end{enumerate}

The first four steps above have to be performed in preparation before the annotation commences. They are ran on yearly basis corresponding to the CWE Top 25. The last step of the two annotation rounds, followed by a third for conflict records, is the bulk of the annotation effort. It is important to note that due to the demanding human expertise and effort, many previous efforts opted for automatic (machine) labelling \cite{AutomaticLabeling,Stucco,GasmiLSTM,STIXnet}; in contrast, we have a high-quality manual effort.  

The resulting dataset is divided into two columns. The first is the CVE id, while the second is a label(s) column. The labels are either single weaknesses, multiple weakness with relationship(one leads/causes the other), or multiple unrelated labels(not encountered in the released dataset). Table~\ref{tab:cveCweDBsample} demonstrates some examples of the dataset. The annotator chooses a number or a group of numbers from 1-25 (corresponding to the CWE rank in Figure~\ref{fig:top25}).  Some records are  simply a label, e.g., 4 which is Improper Input Validation, while others are a causation, e.g., 2-25 (Cross-site Scripting causes Code Injection). 

Moreover, Table~\ref{DBCount} shows the breakdown of single-label records used for the ranking task (excluding multiple labels with relationship). The final annotation process resulted in 4,012 CVE records, 3,605 of which have single labels, and 407 have multiple labels with a causal relationship between them. All labels represent the 2022 CWE Top 25. The table is sorted by the CVSS rank as in Figure~\ref{fig:top25}.

\section{Methodology}
\label{sec:meth}
\begin{figure*}[!htpb]
\centering
    \includegraphics[width=.95\linewidth]{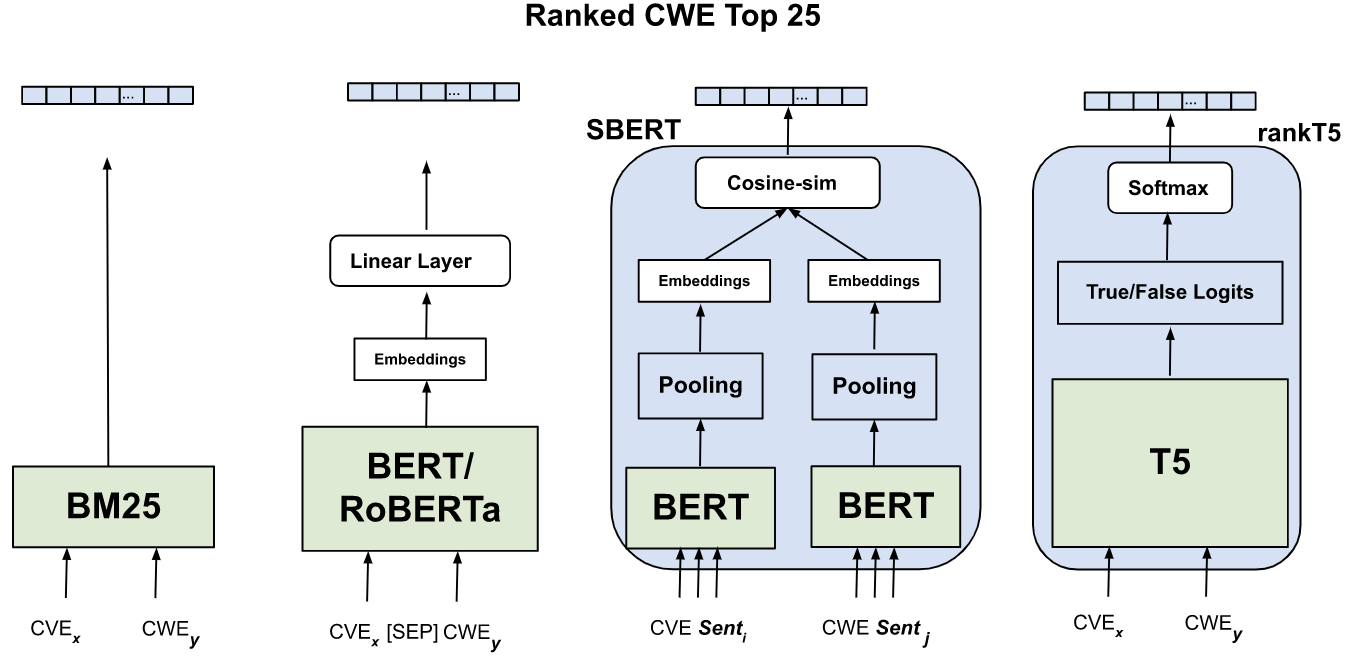}
    \caption{Different model configurations for our ranking task.  }
    \label{fig:allmodels}
\end{figure*}

Our aim is to automate the overall process of mapping vulnerabilities to a given attack framework. Based on the annotated data, the task can be divided into two subtasks: some records entail a single label, and for them, a ranking model suffices, but others imply a relation, a causation of one weakness leading to another one, and as such require an additional structured output step. The overall process is essentially a pipeline, where first the ranking picks the most likely label (document) for a given CVE (query), followed by the second method to induce relationship between the identified weaknesses, if present in the text. Ranking reflects the subjective reasoning amongst annotators. Additionally, recent advances in reinforced learning for human feedback (RLHF) \cite{RLHF} could tremendously benefit the pipeline by integrating the annotation script directly with the models. In this paper, we implement the dataset annotation and the training/inference of the ranking method, and we leave exploring the structured output model and RLHF for future work (the dashed lines in Figure~\ref{fig:pipeline}). Five methods were explored for ranking on our dataset namely BM25, BERT, RoBERTa, SBERT, and T5 models. Figure \ref{fig:allmodels} highlights the application of those models to the ranking task.

\subsection{Pre-Processing}
\label{SubSecPreproc}
Vulnerabilities carry a lot of software- and hardware-specific details including names, versions, URL, emails, CVE ids and  other granular details, which are not necessarily useful at the desired CWE representation. For example, a Cross-site Scripting or an SQL Injection vulnerability present in a WordPress Plugin ought to produce the same model representation as one present in, say, Microsoft SharePoint Foundation. While the above granular details are important for solving the particular issue for that product, the aim here is to produce higher-level representation of the data at the CWE list level. A gazetteer lookup is used for common software/hardware products or packages, e.g., by IBM and Microsoft. Other elements such as domains, emails, and URL are extracted using regular expressions (regex). The cleanup step removes gazetteer entries and regex results, thus reducing the noisy words especially for models relying on word frequencies.   

\subsection{BM25}

BM25 is a popular IR algorithm, which uses TF.IDF and also accounts for the document length and term saturation. It is the baseline model used in this paper. The algorithm retrieves the most relevant document for a given query by generating a score for each query--document pair. In its application to the task, input CVEs are treated as queries and the 2022 CWE Top 25 are considered documents. BM25 is used to retrieve the CWE weakness with the highest relevance, best describing the issue. 
CWE input refers to the concatenation of the name, the description, and the extended description. The CVE input data refer to the description and the title only since the name itself is the id used for indexing solely. As a pre-processing step, we remove the stopwords. Moreover, considering that BM25 relies on the term frequencies, segmenting CWE to sentences, as used in the SBERT similarity measure below, would not improve its performance. 
The outcome of the BM25 model is an array of 25 scores for each CVE record, which we sort to produce a ranking.

\subsection{BERT and RoBERTa}
BERT\cite{BERT} model was the leading breakthrough behind pre-trained language models through masking input sentences. The model is pre-trained to predict the masked token as well as the next sentence(next sentence prediction - NSP). To apply it as a ranking model, we had to fine-tune it on similar textual similarity (STS) as per the BERT guidelines. BERT produces embeddings, i.e.,~vectorial representations of the words, of the CVE query and CWE document pair, and a linear layer is applied to the output logits to rank the labels. Essentially, each query--document pair is encoded, forward-propagated and optimized against the cross-entropy loss of a specific true label. The input data is the tuple $(CVE_x,CWE_y)$, separated by the $[SEP]$ token, such that $x$ enumerates the CVEs in the database and $y$ that of the 25 CWE both at document-level. At inference time, each CVE record is compared to all 25 CWE, choosing the one with the highest output logit.

RoBERTa \cite{Liu-etal-Roberta} aims at improved performance over BERT by training the model for more epochs, using bigger batches, over more data and dynamically changing the mask pattern for the same architecture. As it uses essentially the same architecture as BERT, we train it in a similar fashion.

\subsection{SBERT}
The model is based on sentence-BERT (SBERT) similarity measures \cite{Reimers-SBERT}, which specifically targets the STS task. As demonstrated in previous research~\cite{TCEnet}, document-level models suffer from the loss of details, which affects accuracy. SBERT, however, has been optimized to treat text at the sentence-level, and not document-level, yielding better results~\cite{sbert-net} than BERT. CWE input is segmented into sentences and the model computes the cosine similarity between two sentence embeddings. It was observed that CVE records, in the released dataset, have on average 3.69 sentences. Alternatively, the 25 CWE inputs is a collate of the name, the description, and the extended description with an average of 8.2 sentences. SBERT works by finding the highest cosine similarity between sentences($sent_i$) of the segmented CVE record and sentences of the segmented collage CWE input($sent_j)$. This is achieved through mean pooling over $(sent_i,sent_j)$ embedding combinations. Similar to BM25 model, the cosine similarity yields 25 scores, which we sort to obtain a ranking. The difference between these models is that BM25 is a probabilistic relevance model, based on term frequencies, which focuses on exact textual matching whereas SBERT/BERT/RoBERTa tries to achieve semantic matching.

Pre-trained models are typically fine-tuned on a specific task and a specific dataset, textual ranking and our dataset in this case. Moreover, we further fine-tune the SBERT on our dataset where sentence pair of CVE and CWE are assigned a score $(sent_i,sent_j,score)$. A score of 1.0 is given to the annotated label and 0.0 for a negative sample (CWE labels unlikely to be associated with the CVE record). A random one or two negative samples were used for each positive example.

\begin{table*}
\centering
\begin{tabular}{lccccccccc}
\toprule
 \textbf{Model} & \textit{\textbf{MRR}} & \multicolumn{4}{c}{\textit{\textbf{MAP}@k}} & \multicolumn{4}{c}{\textit{\textbf{NDCG}@k}} \\
  &  & 1& 2 & 3 & 5 & 1& 2 & 3 & 5  \\
\midrule
SBERT (fine-tuned) & \textbf{.9142} & \textbf{.8500} & \textbf{.9057} & \textbf{.9117} & \textbf{.9132} & \textbf{.8446} & \textbf{.9217} & \textbf{.9307} & \textbf{.9334} \\ 
SBERT & .6310 & .4806 & .5653 & .5912 & .6104 & .4778 & .5864 & .6267 & .6582 \\ 
\midrule
rankT5 (fine-tuned) & .8155 & .7115 & .7829 & .7996 & .8104 & .7115 & .8016 & .8266 & .8464  \\
rankT5 & .5570 & .4300 & .4854 & .5081 & .5318 & .4300 & .4999 & .5449 & .5767  \\
\midrule
RoBERTa (fine-tuned) & .2966 & .0693 & .2212 & .2433 & .2593 & .0693 & .2610 & .2957 & .3226 \\
\midrule
BERT (fine-tuned) & .3005 & .0610 & .2240 &  .2416 & .2602 & .0610 & .2667 & .2930 & .3267 \\
\midrule
Baseline BM25 & .1514 & .0166 & .0333 & .0573 & .0860 & .0166 & .0376 & .0737 & .1259 \\ 
\bottomrule
\end{tabular}
\caption{\label{tab:evalres}Evaluation results.}
\end{table*}

\subsection{T5 and RankT5}
T5~\cite{t5Raffel} has shown remarkable performance for many NLP tasks over other deep learning model due to its ability to store more hidden representation for a given task. SBERT, similar to its parent model BERT, uses an encoder-only architecture with a pooling layer, whereas T5 is an encoder-decoder architecture. The approach to ranking is to transform T5 into a token generation problem. Here we use pre-trained and fine-tuned rankT5 \cite{monoT5} models. The model is fine-tuned to predict \emph{true} or \emph{false} as the target token for a relevant or irrelevant query--document pair, respectively,  and then during inference ranking scores are derived. The model was trained by converting the records in the dataset to the required ($query,document$) input and ($true/false$) output. A random one or two $false$ examples for each $true$ example was used.  The existing research revolving T5 does not segment the input into sentences as is the case with SBERT. This is probably due to the ability of transformers to process self-attention and masked-attention present in text in comparison to SBERT's only self-attention. Moreover, T5 was used as a Seq2Seq generating possible labels. Despite not related to the ranking task, the generated text representing possible CWE weaknesses was remarkably human-level after fine-tuning the model. 

\subsection{Training}
All deep learning models were trained by injecting negative examples in the training set. In line with the fact that no two annotators can annotate a list of a few hundred CVEs with the exact CWE labels, the concept of unrelated query--document pair is important to materialize during training. The models learn that those "negative" CWE are unlikely to describe the vulnerability. They are drawn at random from the pool of other 25 CWE labels not chosen by the annotators for that CVE record. Depending on the architecture of the model, this entails either using a zero score, \emph{false} or reserved negative class for SBERT, T5 or BERT/RoBERTa respectively. Moreover, considering the uneven distribution of labels, Table~\ref{DBCount}, the dataset was split into training and validation in a stratified fashion. All models generate a final ranked output of 25 CWE scores, explicitly in the case of BM25 and SBERT, and implicitly as normalized logits in the case of BERT,  RoBERTa and rankT5. 

\section{Experiments}
\label{sec:expNeval}

Ranking tasks are measured using mean reciprocal rank ($MRR$), mean average precision ($MAP$), and Normalized discounted cumulative gain ($NDCG$) \cite{murphy2013machine}. $MRR$ is the inverse of the rank of the first relevant document for each query averaged over all the queries. $MAP$ is the average precision over all the queries, where the average precision is the number of relevant documents in the top-$k$ positions divided by $k$. Top-$k$ represents the first $k$ items of the ordered documents output. $NDCG$ is used if relevance labels have multiple scores. While this is not the case here, the metrics still discounts items later in the list (using $1/log$) giving a slightly different measure than $MAP$ in our case. 
For our ranking task, the actual label is a single value (in our sampled database but could be few labels), whereas the output of the model is an ordered list of all labels. Assuming a model ranks a list of unique labels $ \hat{\mathcal{Y}} = (\hat{y_2},\dots,\hat{y_1},\hat{y_n}) $,  $MAP@k/NDCG@k$ compares the top-$k$ prediction to the ground truth (true positives) $\mathcal{Y}=y_{true}$.  

Typical values for $k$ in the ranking literature are 5 or 10.  Expectedly, the higher the value of $k$ is, the higher accuracy is achieved since it covers more of the ordered list, thus increasing the likelihood of finding the relevant document. Considering that $\approx$ 89\% of rows have single labels, we opted for choosing more concise values of $k={1,2,3,5}$, with $MAP@1$ and $NDCG@1$ representing a classification measure.

All experiments were conducted on NVIDIA's Quadro RTX 6000 with 25\textit{GB} of RAM using Python3.7+, and CUDA library. The pre-trained deep learning models were leveraged from HuggingFace. 

\begin{table}
\centering
\resizebox{0.5 \textwidth}{!}
{
\begin{tabular}{@{}lcccc@{}}
\toprule
\bf Metric & \multicolumn{2}{c}{\bf BM25}  & \multicolumn{2}{c}{\bf SBERT}  \\
& \bf  -preproc. & \bf +preproc. & \bf  -preproc. & \bf +preproc. \\
\midrule
\textit{MRR} &  .1493 & \textbf{.1514} & .6272 & .6310  \\
\textit{MAP@1} & .0166   & \textbf{.0263} & .4833 & .4806   \\ 
\textit{MAP@2} & \textbf{.0374} & .0333   & .5583 & .5653   \\
\textit{MAP@3} &  .0532 & \textbf{.0573} & .5861 & .5912   \\
\textit{MAP@5} &  .0784 & \textbf{.0860} & .6068 & .6104   \\
\bottomrule
\end{tabular}
}
\caption{\label{tab:preprocRes} Effect of pre-processing on BM25 scores.}
\end{table}

\begin{table*}[t]
\renewcommand{\arraystretch}{1.1}
\centering
\resizebox{1 \textwidth}{!}
{
\begin{tabular}{@{}ccp{2.9cm}p{2.9cm}p{2.9cm}p{2.9cm}@{}} 
\toprule
& \bf Hyperparameters & \bf{SBERT} & \bf{rankT5} & \bf{BERT} & \bf{RoBERTa} \\
\hline
\multirow{11}{*}{Pretrained} &  Model & all-mpnet-base-v2 & monot5-base-msmarco-10k & bert-base-uncased & roberta-base  \\
 & Implementation & HuggingFace$^*$  &   HuggingFace$^+$ & HuggingFace$^{\bullet}$ &HuggingFace$^{\bowtie}$  \\
 & Dimensions &  768 & 768 & 768 & 768  \\ 
 & Max-Seq length & 384 & 512 & 512 & 512 \\
 & Optimizer & AdamW & AdamW & AdamW & AdamW \\
 & Precision & 32 bits & 32 bits & 32 bits & 32 bits \\ 
 & Size & 420 MB & 851 MB & 418 MB & 476 MB\\ 
 & Suitable usage & Cosine similarity & Text generation & Next sent. prediction & Next sent. prediction \\
  & Pretrained data & 1B+ training pairs- general purpose data &  8.8M passage - MS-MARCO & BookCorpus and Eng. Wikipedia & Five datasets$^{\bowtie}$ of 160GB of text  \\
\cline{2-6}
 &Tokenizer &  	microsoft/mpnet-base & t5-base & bert-uncased & roberta-base \\
  &Vocab size & 30527 & 32128 & 30522 & 50265 \\
\bottomrule
\multirow{5}{*}{Fine-tuned} & Epochs  & 10 & 10 & 5 & 20\\
& Batch size & 4 & 2 & 8 & 32 \\
& Learning rate & 2$e-5$ & 3$e-4$ & 2$e-5$ & 2$e-5$\\
& Warmup steps & 2884 & 1000 & 0 & 0  \\
& Weight decay & 0.01 & 5$e-5$ & 0.01 & 0.01 \\
\bottomrule
\end{tabular}
}
\footnotesize{$^*$\url{ https://huggingface.co/sentence-transformers/all-mpnet-base-v2}} \\
\footnotesize{$^+$\url{https://huggingface.co/castorini/monot5-base-msmarco-10k}} \\
\footnotesize{$^{\bullet}$\url{https://huggingface.co/bert-base-uncased}} \\
\footnotesize{$^{\bowtie}$\url{https://huggingface.co/roberta-base}} \\
\caption{\label{tab:hyperparam}Deep learning models' settings.}
\end{table*}

\subsection{Pre-Processing}
The pre-processing step, as described in Sub-Section \ref{SubSecPreproc}, attenuates the noise due to nuance pieces of information like software versions. The results show that BM25 is the prime beneficiary of such a cleanup as shown in Table~\ref{tab:preprocRes}. For comparison, we also demonstrate the same pre-processing on the pre-trained SBERT. BM25 improved in all metrics except \textit{MAP@2}, e.g.,~it increased by 1.4$\%$ and 9.7$\%$ for the \textit{MRR} and the \textit{MAP@{5}} measures, respectively, with an average improvement of 13.3\% across all metrics.  Pre-trained SBERT, on the other hand, barely improved by 0.61$\%$ and 0.59$\%$ for the same \textit{MRR} and \textit{MAP@5} metrics with an average improvement of 0.55\% across all metrics. The improvement for BM25 is due to its reliance on term frequencies, which benefit more from the removal of noisy words, SBERT on the other hand uses more semantic matching and the encoder architecture learns to pay less attention to those noisy words.

\subsection{Results}
The simplicity of the BM25 IR model is that the CWE labels are treated as a corpus of documents directly without the need to fine-tune. BM25 scores reflect the use of exact matching, which resulted in the least $MRR$, $MAP@k$ and $NDCG@k$ values across all other models as seen in Table~\ref{tab:evalres}. BERT and RoBERTa improved over BM25 due to their similarity semantic methodology, using STS fine-tuning. The pre-trained BERT (original release) was used for NSP task and needed fine-tuning to produce the results. Both models had very comparable results considering that they share the same architecture, but the pre-trained RoBERTa was trained on a larger corpus, for longer epochs and with larger batches. In order to produce better RoBERTa results, one had to perform the proposed exploratory exercises~\cite{Liu-etal-Roberta}. RoBERTa ran using a batch of 32, due to limited memory available(25GB GPU), where as the original paper~\cite{Liu-etal-Roberta} saw improvements at 256+ and even 8K batch sizes. Moreover, RoBERTa slightly outperforms BERT noticeably on $MAP/NDCG@1$  by 13.63\% (and slightly on $MAP/NDCG@3$ by 1.14\% / 0.9\%), whereas BERT generalized better on other metrics by an average of 1.2\%. This supports the view that RoBERTa could have produced better results if it was fine-tuned longer using bigger batches.

SBERT operates at the sentence level and its mean pooling returns the highest cosine scores for each \textit{(CVE record-CWE label)} pair. Ranking on this dataset is in need of deep contextual understanding especially that they share many common words, for example, Missing Authorization (CWE-862) and Improper Authentication (CWE-287) threats are often described similarly as a lack of proper login and access control to certain modules in the software/hardware stack. This by itself is a clear advantage of SBERT over BM25. Moreover, SBERT clearly outperformed RoBERTa and BERT due to its highly specialized architecture built to measure the sentence similarity and at the granular sentence level vs. the document level which BERT/RoBERTa operates at. 

Moreover, a pre-trained SBERT achieved noticeably higher accuracy, outperforming a pre-trained rankT5. This is owed to the sheer amount of data (one billion sentence pairs) that the SBERT model was trained on (see Table~\ref{tab:hyperparam}). Despite the view that T5 is more capable at deducing deep representations, this proves how ample training data propels the performance of a less-complex yet specialized model. 
Our experiments also demonstrate the importance of fine-tuning deep learning models on custom data. rankT5 performance increased by 56.3\% on average across all metrics; similarly, SBERT improved by 56.7\% across all metrics.  Both rankT5 and SBERT were fine-tuned on our dataset using the same number of 10 epochs. BERT was trained using the recommended 4 or 5 in the original implementation, while RoBERTa could always benefit from longer training and larger batch sizes.  
Table~\ref{tab:hyperparam} shows the values of the hyper-parameters of the pre-trained and fine-tuned deep learning models in our code repository\footnote{\url{https://github.com/ahadda5/CVE-NLP}}.
Both pre-trained and fine-tuned models were validated on a 20\% test set (80-20\% training/test split) using the same random seed throughout all experiments. Furthermore, a fine-tuned SBERT, being pre-trained on a huge corpus and custom built for the task of STS scored the highest across all metrics compared to all models. We believe that a ranking T5 pre-trained on the same 1B+ pairs, as the pre-trained SBERT, and further fine-tuned on our dataset would have comparable performance to SBERT. 
Moreover, none of the used models were custom-built for this cyber-security dataset, but rather pre-trained models fine-tuned on the dataset. We believe that a custom-built encoder architecture with calibrated dropout, mean-pooling and linear layers can achieve results comparable to SBERT/rankT5 for a fraction of the parameters(size). 

\begin{table}[ht]
\footnotesize
\centering
\begin{tabular}{lp{5.8cm}r}
\toprule
\textbf{Id} & \textbf{Generated Label} & \textbf{F1}  \\
\hline
CWE-787 & Out-of-bounds Write & 0.69 \\
CWE-79 & Improper Neutralization of Input During Web Page Generation (Cross-site Scripting)  & 0.96 \\
CWE-89 & Improper Neutralization of Special Elements used in an SQL Command  (SQL Injection) & 0.99  \\
CWE-20 & Improper Input Validation & 0.64 \\
CWE-125 & Out-of-bounds Read & 0.92  \\
CWE-78 & Improper Neutralization of Special Elements used in an OS Command (OS Command Injection) & 0.48  \\
CWE-416 & Use After Free &  0.92 \\
CWE-22 & Improper Limitation of a Pathname to a Restricted Directory (Path Traversal) & 0.75 \\
CWE-352 & Cross-Site Request Forgery (CSRF) & 0.88 \\
CWE-434 & Unrestricted Upload of File with Dangerous Type & 0.71 \\
CWE-476 & NULL Pointer Dereference & 0.62 \\
CWE-502 & Deserialization of Untrusted Data & 0.84 \\
CWE-190 & Integer Overflow or Wraparound & 0.75  \\
CWE-287 & Improper Authentication &  0.63  \\
CWE-798 & Use of Hard-coded Credentials & 0.31 \\
CWE-862 & Missing Authorization & 0.55  \\
CWE-77 & Improper Neutralization of Special Elements used in a Command (Command Injection) & 0.65 \\
 CWE-119 & Improper Restriction of Operations within the Bounds of a Memory Buffer & 0.26 \\
 CWE-276 & Incorrect Default Permissions & 0  \\
 CWE-918 & Server-Side Request Forgery (SSRF) & 0.62 \\
 CWE-362 & Concurrent Execution using Shared Resource with Improper Synchronization (Race Condition) & 0.75  \\
 CWE-400 & Uncontrolled Resource Consumption & 0.65 \\
 CWE-611 & Improper Restriction of XML External Entity Reference & 0.67 \\
 CWE-94 & Improper Control of Generation of Code (Code Injection) & 0.55 \\
\midrule
\textbf{\textit{New}} &   double-free  &  0 \\
\textbf{\textit{New}} &   floating point  &  0 \\
\textbf{\textit{New}} &   improper interference with a pathname to a restricted directory (path traversal)  &  0 \\
\textbf{\textit{New}} &  instructor-led initiative   &  0 \\
\textbf{\textit{New}} &   unencrypted data  &   0 \\
\textbf{\textit{New}} &   upload of file with dangerous type &   0 \\
\textbf{\textit{New}} &    use count &   0 \\
\midrule 
 &  \multicolumn{1}{r}{Macro avg F1} & \multicolumn{1}{l}{\textbf{0.51}} \\
 \bottomrule
\end{tabular}
\caption{\label{t5genlabels}
T5-generated weaknesses on the test set.}
\end{table}

Lastly, due to the versatile design of T5, which handles many NLP tasks, it was fine-tuned to generate weaknesses rather than rank them. While this is not part of our pipeline, it merits the experimentation. To that effect, Table~\ref{t5genlabels} shows the results. The model generated 31 weaknesses on the same 20\% split. The validation set had all weaknesses excluding CWE-306, i.e.,~24 labels were present. $F1$ score is used for generative output and the model demonstrated a macro-average $F1$ of 0.51. High accuracy was achieved on frequent threats like 0.96 on \emph{Cross-site Scripting} vulnerability. The 7 newly generated weaknesses had zero $F1$ accuracy for not being present in the CWE labels. However, they were still remarkably meaningful, demonstrating the generative capability of the latest LM (Language Model).

\section{Conclusion and Future Work}
\label{sec:concNFut}
In general, AI researchers tend to emphasize model designs more so than training and validation datasets, leading to the creation of synthesized data just to validate their designs. This challenge in the field of automation of cyber-security analysis and reporting is a huge hurdle. The general vision of this paper is to help the cyber-security community achieve automatic mapping of CVE records to a structured format(e.g.,~MITRE CWE Top 25). In order to do that, an open-source dataset was needed, and is one of the main contributions of our work. The same annotation approach can be used for formats such as the MITRE ATT\&CK framework. Moreover, models with the ability to handle the very nature of annotation and the varying annotators' opinions were evaluated. A classification approach is insufficient when multiple cyber-security analysts might occasionally disagree on the underlying weaknesses. A ranking task is more appropriate particularly when weaknesses could infer a chain of relationships between them. In this paper, a pre-trained SBERT and a ranking T5 model achieved high accuracy. A pre-trained SBERT on a huge corpus enabled us to beat T5's performance, despite the encoder and decoder enhanced T5 representations in comparison to SBERT's encoder-only sentence embeddings. Furthermore, fine-tuning both models on our dataset brought about considerable improvements achieving results in line with the latest research in other domains(MS-MACRO~\cite{msmarco} and NQ~\cite{NQ}) \cite{rankT5google}. SBERT greatly exceeded BM25 accuracy due to the deep contextual understanding required and not simply reliance on word matching. BERT and RoBERTa did not fair well with a ranking task based on textual similarly the same way SBERT or rankT5 did, which are built for STS. SBERT specially built to infer similarity between two sentences, in our case CVE and CWE sentences, achieved the highest amongst all the models.

\vspace{.5cm}
There are two interesting directions for future work:

\begin{itemize}
\item Recent advances in human prompts would allow our annotation script to be integrated with the ranking model to allow reinforced learning for human feedback (RLHF) similar to the current OpenAI chat models \cite{RLHF} (e.g.,~ChatGPT). The ranking model would benefit from a majority vote of annotators on CVE records allowing it to continuously fine-tune its scoring. This human-in-the-loop design, as shown in Figure~\ref{fig:pipeline}, would be a great tool for the cyber-security community.
\item  A model has to be explored to determine the causation between weaknesses, if present, thus producing the final structured output.  Currently, there are 407 dataset records that require that extra step in the pipeline, which we did not explore here. Unlike, the ranking task which requires examples of unrelated query--document pairs added to the training set, this step has the existing 3,605 single-labels to train the model via negativa. 
\end{itemize}

\section*{Ethics and Broader Impact}

We do not expect any harm coming from this research. In fact, the purpose is quite the opposite: to assist the cyber-security community with tools to identify threats. 

\textbf{Annotation}. The annotation was conducted by cyber-security experts running the provided script. Several meetings were conducted to make sure the approach is consistent. The experts are working for a cyber-security research institution, not affiliated with MITRE or any organization mentioned in this paper, and were duly enumerated as part of their daily jobs in agreement with the authors. 

\textbf{Biases}. We acknowledge that the annotation carried out by the group of cyber-security experts might carry certain biases. However, the released script allows users to alter the labels they may disagree with. 

\textbf{Environmental Impact}. The released dataset of 4,012 records is quite small and allows for training in a rather short period of time (2-3 hours for SBERT/T5). We warn, however, if such models are trained at the full scale of the NVD more computational power and time on GPUs is needed which may contribute to global warming~\cite{strubell-etal-2019-energy}. 
\bibliography{refs}
\bibliographystyle{plain}


\end{document}